\documentclass[twocolumn,showpacs,pra,floatfix,aps,reprint]{revtex4-1}

\usepackage[latin1]{inputenc}
\usepackage[american]{babel}
\usepackage{amsmath}
\usepackage{amssymb}
\usepackage{latexsym}
\usepackage{amsfonts}
\usepackage{dsfont}
\usepackage{graphicx}
\usepackage{bm}
\usepackage{natbib}

\newcommand{\num}{n}

\newcommand{\mode}[1]{\ensuremath{\mc{#1}}}
\newcommand{\ket}[1]{\ensuremath{|#1\rangle}}
\newcommand{\bra}[1]{\ensuremath{\langle #1 |}}

\newcommand{\mf}[1]{\boldsymbol{#1}}

\newcommand{\eps}{\epsilon}
\newcommand{\vro}{\varrho}

\newcommand{\be}{\begin{equation}}
\newcommand{\ee}{\end{equation}}
\newcommand{\ba}{\begin{eqnarray}}
\newcommand{\ea}{\end{eqnarray}}
\newcommand{\bali}{\begin{align}}
\newcommand{\eali}{\end {align}}

\newcommand{\mc}[1]{\ensuremath{\mathcal{#1}}}
\newcommand{\ix}[1]{\text{ #1}}
\newcommand{\bc}{\begin{center}}
\newcommand{\ec}{\end{center}}
\newcommand{\bi}{\begin{itemize}}
\newcommand{\ei}{\end{itemize}}
\newcommand{\mean}[1]{\ensuremath{ \langle\,#1\, \rangle}}

\newcommand{\foc}[1]{\ensuremath{\mc T_{F_{#1}}}}
\newcommand{\so}[1]{\ensuremath{\mf g^{(2)}_{F_{#1}}}}

\newcommand{\Sp}[1]{\mathcal{S}_{#1 }^{\,+}}
\newcommand{\Sm}[1]{\mathcal{S}_{#1 }^{\,-}}

\newcommand{\rhoA}{\vro_{\text{atom}}}
\newcommand{\drhoA}{\del{t}\vro_{\text{atom}}}
\newcommand{\HA}{\mc H_{\text{atom}}}
\newcommand{\HAB}{\mc H_{\mc{AB}}}

\newcommand{\del}[1]{\partial_{#1}}
\newcommand{\rhoexp}{\rho_{exp}}

\newcommand{\pop}{\mc P}

\begin{document}

\title{Microcavities coupled to multilevel atoms}

\author{Sandra Isabelle \surname{Schmid}}
\email{sandra.schmid@mpi-hd.mpg.de} 

\author{J\"org \surname{Evers}}
\email{joerg.evers@mpi-hd.mpg.de} 

\affiliation{Max-Planck-Institut f\"ur Kernphysik, Saupfercheckweg 1, D-69117 Heidelberg, Germany} 

\date{\today}

\allowdisplaybreaks[3]

\begin{abstract}
A three-level atom in the $\Lambda$-configuration coupled to a microcavity is studied. The two transitions of the atom are assumed couple to different counterpropagating mode pairs in the cavity. We analyze the dynamics both, in the strong-coupling and the bad cavity limit. We find that compared to a two-level setup, the third atomic state and the additional control field modes crucially modify the system dynamics and enable more advanced control schemes. All results are explained using appropriate dressed state and eigenmode representations. As potential applications,  we discuss optical switching and turnstile operations and detection of particles close to the resonator surface.
\end{abstract}


\maketitle


\section{Introduction}
Whispering gallery microcavities are an attractive implementation for quantum optical setups, often motivated by the very low loss rate that can be achieved. This has led to a multitude of applications~\cite{N5,review}, such as  switches~\cite{Rao2010,cleopop2008-old,popovic1,switch2010,cascade2009}, transistors ~\cite{Hong2008a,Xiong2008}, quantum networks ~\cite{Yao2005,quantint}, photonic entanglement~\cite{Ajiki2006}, or optomechanics~\cite{ISI:000289199400038, ISI:000268670300035,ISI:000276205000034,Weis10122010,schwab}.

The light is confined to the resonator by total internal reflection, which leads to an exponentially damped evanescent field around the cavity. On the one hand this allows to couple the cavity, e.g., to an external fiber. But on the other hand, the evanescent field also allows the resonator field to interact with nearby particles. Since the cavity volume and the resonator linewidth can be small, even the near-resonant coupling of a single atom can have a substantial effect on the resonator dynamics~\cite{Aoki2009} and modify the photon statistics of transmitted photons~\cite{Ajiki2008}. In~\cite{turnstile}, a photon turnstile was created by coupling a two level atom to the cavity. The sensitivity of the resonator to nearby atoms can also be used to measure the presence of nearby particles, their optical properties, or their concentration~\cite{particledet2010,detvirus2011,vollmer2008,unserloop,vollmer2, armani2007, arnold10}.
The evanescent field can also be used to trap atoms, as has been demonstrated for atoms coupled to a tapered fiber~\cite{PhysRevA.70.063403,Dowling,PhysRevLett.104.203603}.

In most cases, the particle interacting with the resonator is treated as a two-level atom, but there is some related previous work on more complex level schemes. For example, in~\cite{Hong2008a} a three level atom in $\Lambda$ configuration coupled to a nearby microcavity was considered. Here, however, the atom only couples to the cavity with one of its transitions, while the other one is coupled to an external laser field in order to control the atom-resonator interaction. In~\cite{Ajiki2009}, a three level $V$ system inside a cavity was considered to create entangled photon pairs.  In~\cite{Milow}, a three-level atom in $V$-configuration coupled to a resonator is studied, focusing on the effect of the  scattering between the different cavity modes on the optical spectra. Transmission and reflection amplitudes for single photon wave packets interacting with a resonator coupled to a three-level atom are considered in~\cite{Cheng}. Finally, in a related work~\cite{1367-2630-12-4-043052}, photon scattering from
a three-level atom coupled to a 1D wave guide was studied.
These works are not least motivated by the fact that a setup  in which several transitions of the particle are coupled to respective modes of a single resonator, potentially could be more flexible and could enable more advanced control schemes.

\begin{figure}[b]
\centering
\includegraphics[width=8.5cm]{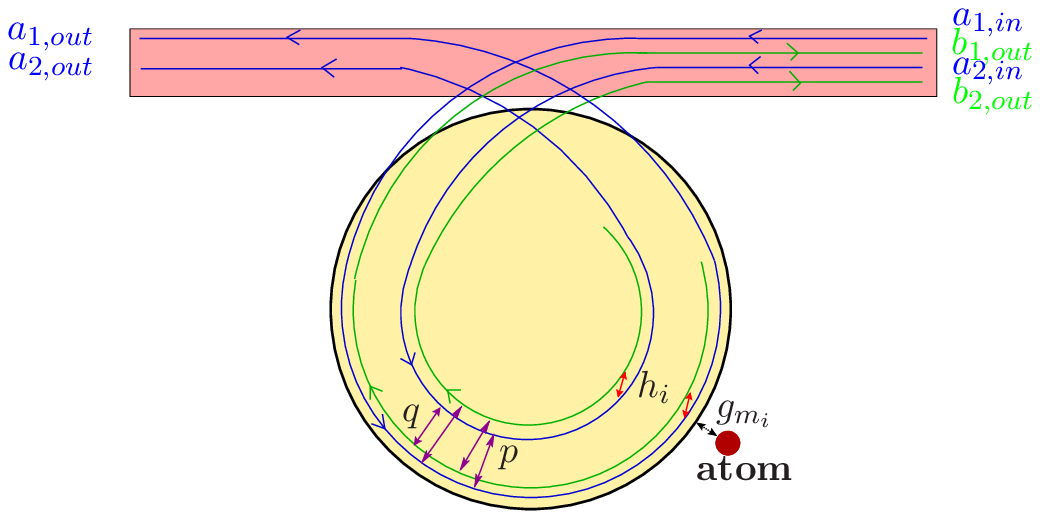}
\includegraphics[width=4cm]{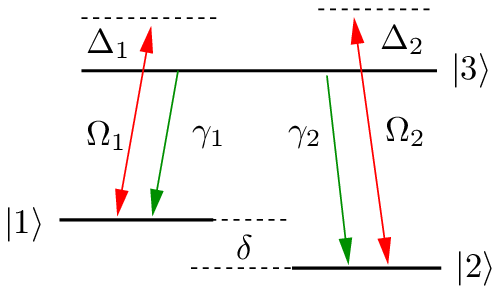}
\caption{\label{sys}(Color online) Schematic setup. The upper panel shows the resonator coupled to the fiber and an atom. The lower panel shows the internal structure of the $\Lambda$-type atom.}
\end{figure}

\begin{figure}[t]
\centering
\includegraphics[width=8cm]{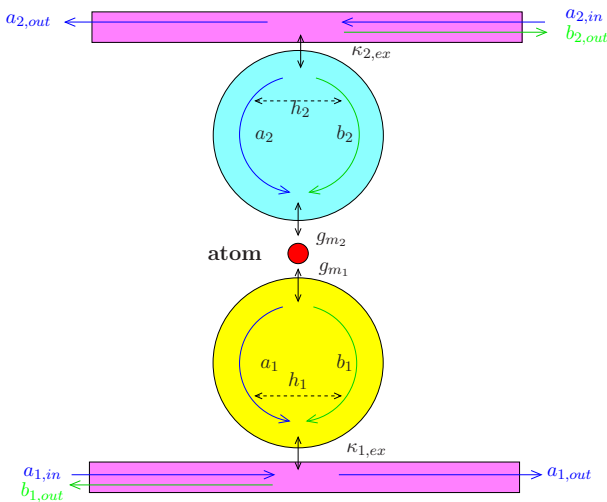}
\caption{\label{twocav}(Color online) Alternative implementation of our model system based on two microcavities. Without the atom, the resonators do not interact, as they are not in resonance. The atom in $\Lambda$-configuration acts as a quantum link, with the two transitions each of which coupling to one of the resonators, respectively.}
\end{figure}

To demonstrate this, here, we analyze advanced control schemes in the generic model system of an atom coupled to a resonator, probed by a fiber coupled to the resonator. In particular, we study a three-level atom in the $\Lambda$ configuration coupled to a microcavity, see Fig.~\ref{sys}. We consider the case in which each atomic transition couples to a pair of counterpropagating cavity modes, respectively. In addition, the resonator is coupled to a fiber which allows to probe the atom-cavity setup. We analyze the dynamics and the optical properties of the system in dependence on the fields applied via the fiber using two different approximations with complementary validity ranges. First, we assume weak input fields, and restrict the Hilbert space to low excitations.  Second, we adiabatically eliminate the cavity and derive an effective master equation for the atomic degrees of freedom alone. 
We find that  the existence of two coupling mode pairs and two atomic ground states crucially changes the systems dynamics as compared to the two level case and opens up new possibilities to control the coupled system. For this, we typically treat one of the fields as probe field, and the other as control field. As potential applications, we discuss controllable optical switching and  photon turnstile operations, as well as the  measurement of the position of particles close to the resonator. All results are interpreted based on the system's dressed states or eigenmodes.

Our system could be realized, for example, by tuning the Zeeman splitting of two atomic magnetic sublevels to the frequency spacing of two mode pairs in the resonator. Alternatively, different polarizations couple be used to selectively couple the transitions to different modes, but the leakage pattern of evanescent fields with different polarizations around the resonator surface is not uniform. Finally, also a setup with two nearby resonators as shown in Fig.~\ref{twocav} could be considered in which the resonators do not interact in the absence of the atom as they are not in resonance with each other. The atom could then act as a quantum link between the resonators, with each transition coupling to one of the resonators, respectively.

This article is organized as follows. In Sec.~\ref{sysma}, we introduce our model system derive the equations of motion. In Sec.~\ref{twomethods}, we describe the two calculation methods based on a truncation of the Hilbert space (\ref{fock}) or on an adiabatic elimination of the resonator (\ref{gencouplings}) used throughout the analysis in detail. Our observables, the transmission and reflection intensities as well as second order correlations are introduced in Sec.~\ref{observables}. Section~\ref{results} discusses our results. We present our findings for the strong coupling regime in Sec.~\ref{strong} and explain them via a dressed state atomic level scheme. Afterwards we show how our setup in this parameter regime could be used as a precision measurement device for the position of a nearby atom. In Sec.~\ref{badcav} we show our results for the bad cavity regime. We present transmission and reflection intensities as well as our studies on the photon statistics. Finally in Sec.~\ref{apply} we show how our coupled system can be used as a tunable bimodal photon turnstile in the bad cavity regime.


\section{Theoretical considerations}

\subsection{\label{sysma}Model system}

The system we investigate is shown in Fig.~\ref{sys}. It consists of a toroidal microresonator, which is driven by a fiber, and coupled to an atom placed within the evanescent field surrounding the resonator. The atom is modeled with three states in the $\Lambda$-configuration with a energy separation $\delta$ of the two ground states.  The resonator is modeled with two pairs of counterpropagating whispering gallery modes, with one pair coupling to each transition in the atom, respectively. This selective coupling could be achieved either via different mode frequencies, or via different polarizations. An alternative possible realization is shown in Fig.~\ref{twocav}, and will be discussed in more detail in Sec.~\ref{summ}.
In our system a number of different coupling mechanisms are of importance. First, within the resonator, each pair of counterpropagating  modes with photon destruction operators $a_i, b_i$ is coupled via the scattering rate $h_i$ ($i\in\{1,2\}$).  In principle, also photons from one mode pair could be scattered into the other pair of modes. This coupling is described by the parameters $p$ and $q$. But in particular if the two mode pairs have different frequencies or polarizations, $p$ and $q$ will be negligible. 
Next, the fiber couples via the evanescent field to the cavity such that photons of the two probe fields with coupling strength $\mc E_{1}$ and $\mc E_{2}$ can enter the cavity. This coupling is described by the constants $\kappa_{m_i,ex}$, where $m\in\{a,b\}$ and $i\in\{1,2\}$. 
The incoherent decay of the cavity modes into the vacuum we include by the internal decay rate $\kappa_{m_i,in}$. Then, the overall loss rate for the cavity modes is 
\begin{align}
\kappa_{m_i}=\kappa_{m_i,in}+\kappa_{m_i,ex}\,.
\end{align}
Finally, the coupling of the atom to the resonator is described by the constants $g_{m_i}$. The respective coupling strength depends 
on the position of the atom characterized by the radial position $r$ and the azimuthal position $x$ along the resonator circumference, 
and is of the form
\begin{subequations}
\label{gs}
\begin{align}
g_{a_i}&=g_0^i f(r,z) e^{ik_ix}\,,\\
g_{b_i} &= g_{a_i}^*\,.
\end{align}
\end{subequations}
Here, $k_i$ is the wave number of the mode pair $i$ and $z$ is the vertical coordinate. The first mode pair $i=1$ couples to the first atomic transition $\ket{1}\leftrightarrow\ket{3}$,  whereas the second pair $i=2$ couples to the second transition $\ket{2}\leftrightarrow\ket{3}$. We further define the atomic transition operators
\begin{subequations}
\begin{align}
\Sp{i} &= \ket{3}\bra{i}\,,\\
\Sm{i} &= \ket{i}\bra{3}\,.
\end{align}
\end{subequations}
Since in general the frequencies of the driving fields, the resonator modes, and the atomic transition frequencies may be different, we introduce the detunings 
\begin{subequations}
\begin{align}
\eps&=\omega_p^1-\omega_p^2\,,\\
\delta_i&=\omega_c^i-\omega_p^i\,,\\
\Delta_i&=(\omega_3-\omega_i)-\omega_p^i\,,
\end{align}
\end{subequations}
where $\omega_c^i$ are the two cavity mode pair resonance frequencies, $\omega_p^i$ the laser frequencies of the probe fields in the fiber, 
and $\omega_3-\omega_i=\omega_a^i$ the atomic transition frequencies.

With these definitions, the Hamiltonian ${\mc H}_{ab}$ for our system can be written as
\begin{subequations}
\label{ham1}
\begin{align}
{\mc H}_{ab} &= {\mc H}_{ab}^{0} +{\mc H}_{ab}^{s} + {\mc H}_{ab}^{c} \,,\\
{\mc H}_{ab}^{0} &=-\hbar \sum\limits_{i=1}^2 \Delta_i\, \mc{S}_i^-\mc{S}_i^+ +\hbar\sum\limits_{i=1}^2\sum_{m\in{a,b}}\delta_i m_i^{\dagger} m_i \,,\\
{\mc H}_{ab}^{s} &= \hbar p (a_1^\dagger a_2+b_1^\dagger b_2) e^{i\eps t}+ \hbar q (a_1^\dagger b_2+b_1^\dagger a_2) e^{i\eps t} \nonumber \\
&\quad + \hbar \sum_{i=1}^2 h_i a_i^\dagger b_i  + \textrm{H.c.}\,,\\
{\mc H}_{ab}^{c} &= \hbar \sum_{i=1}^2 \mc E^*_{i} a_i  + \hbar \sum\limits_{i=1}^2\sum_{m\in{a,b}}  g_{m_i}^* m_i^\dagger\mc S_i^- + \textrm{H.c.}\,,
\end{align}
\end{subequations}
where ${\mc H}_{ab}^{0}$ contains the free energies of the atom and the cavity, ${\mc H}_{ab}^{s}$ the scattering between the different modes within the resonator, and ${\mc H}_{ab}^{c}$ the coupling between fiber and resonator as well as from resonator to the atom. The subindex $ab$ indicates that the Hamiltonian is written in terms of the resonator modes $a_i$ and $b_i$. Including the incoherent processes leads to the master equation
\begin{align}
\del{t}\rho=&-\frac{i}{\hbar}\left[{\mc H}_{ab},\rho \right] 
 + \mc L_{\kappa}\rho+\mc L_{\gamma}\rho\,,\\
\mc L_{\kappa}\rho =& \sum\limits_{i=1}^2\sum\limits_{m\in\{a,b\}}\,\kappa_{m_i}\left(2 m_i\rho m_i^\dagger-m_i^\dagger m_i \rho \right . \nonumber \\
&\quad \left. -\rho m_i^\dagger m_i \right) \,,\\
\mc L_{\gamma}\rho =& \sum\limits_{i=1}^2\frac{\gamma_i}{2}\left(2\mc S_i^-\rho\mc S_i^+-\mc S_i^+\mc S_i^-\rho-\rho \mc S_i^+\mc S_i^-\right)\,,
\label{cme}
\end{align}
where $\mc L_{\kappa}\rho$ describes cavity decay, and $\mc L_{\gamma}\rho$  the atomic spontaneous emission.

We next introduce the normal modes~\cite{strong}
\begin{subequations}
\label{normalmodes}
\begin{align}
\mc A_i&=\frac{a_i+b_i}{\sqrt{2}}\,,\\
\mc B_i&=\frac{a_i-b_i}{\sqrt{2}}\,,
\end{align}
\end{subequations}
which leads to a partial diagonalization of the Hamiltonian Eqs.~(\ref{ham1}) to give
\begin{subequations}
\begin{align}
{\mc H}_{\mc A \mc B} &= {\mc H}_{\mc A \mc B}^{0} + {\mc H}_{\mc A \mc B}^{c} + {\mc H}_{\mc A \mc B}^{a} + {\mc H}_{\mc A \mc B}^{s}\,,\\
{\mc H}_{\mc A \mc B}^{0}&=-\hbar \sum_{i=1}^2 \Delta_i \Sm{i} \Sp{i} +\hbar \sum_{i=1}^2(\delta_i+h_i) \mc A_i^+\mc A_i \nonumber \\
&\quad +\hbar \sum_{i=1}^2(\delta_i-h_i) \mc B_i^+ \mc B_i \,,\\
{\mc H}_{\mc A \mc B}^{c}&= \frac{\hbar}{\sqrt{2}}  \sum_{i=1}^2 \left(\mc E_i^*\mc A_i+\mc E_i^*\mc B_i \right) +  \ix{H.c.}\,,\\
{\mc H}_{\mc A \mc B}^{a} &= \hbar  \sum_{i=1}^2 \left (g_{\mc A_i} \, \mc A_i^+ \, \Sm{i}  -i  g_{\mc B_i} \,  \mc B_i^+ \, \Sm{i}\right ) + \textrm{H.c.}\,,\\
{\mc H}_{\mc A \mc B}^{s} &= \hbar (p-q) e^{i\eps t}\mc B_1^+ \mc B_2 \nonumber \\
&\quad + \hbar (q+p) e^{i \eps t}\mc A_1^+ \mc A_2 +\ix{H.c.}\,,
\label{ABham}
\end{align}
\end{subequations}
where $g_{\mc A_i}=g_0^{(i)}\cos(k_ix)$ and $g_{\mc B_i}=g_0^{(i)}\sin(k_ix)$.

Transforming also the incoherent parts to the new basis, we obtain the equations of motion for the normal modes as
\begin{subequations}
\label{modes}
\begin{align}
\del{t} \mc A_i(t)=&-\frac{i}{\hbar}[ \mc A_i(t), {\mc H}_{\mc A \mc B}]-\kappa_{ \mc A_i} \mc A_i \nonumber  \\
=& -i (\delta_i+ h_i)\mc A_i -i \frac{\mc E_i}{\sqrt{2}} -i g_{ \mc A_i}  \Sm{i} \nonumber \\
& - i (q+p)e^{i\epsilon t}\mc A_{\neg i} -\kappa_{ \mc A_i} \mc A_i\,,\\
\del{t}  \mc B_i(t) =& -\frac{i}{\hbar}[ \mc B_i(t),{\mc H}_{\mc A \mc B}]-\kappa_{ \mc B_i} \mc B_i\nonumber\\
=& -i(\delta_i-h_i)\mc B_i -i \frac{\mc E_i}{\sqrt{2}}- g_{ \mc B_i} \Sm{i} \nonumber \\
&-i(p-q)e^{i\epsilon t}\mc B_{\neg i}  -\kappa_{ \mc B_i} \mc B_i\,,
\label{ABeom}
\end{align}
\end{subequations}
where the two mode pairs are labeled by $i\in\{1,2\}$, and $\neg i$ is 1 for $i=2$ and vice versa.
Since modes of one pair $\{a_i,b_i\}$ are of the same frequency, it is reasonable to assume $\kappa_{a_i}=\kappa_{b_i}=\kappa_i$ holds. The incoherent part remains diagonal also after the basis transformation, and we can write $\kappa_{\mc A_i}=\kappa_{\mc B_i}=\kappa_i$ and analogously for the internal and external decay rates $\kappa_{\mc A_i,ex}=\kappa_{\mc B_i,ex}=\kappa_{i,ex}$ and $\kappa_{\mc A_i,in}=\kappa_{\mc B_i,in}=\kappa_{i,in}$, respectively.
As expected, it can be seen from Eqs.~(\ref{modes}) that for $p=0=q$, the different normal modes $\mc A_i$ and $\mc B_i$ become independent apart from the coupling via the atom.

\subsection{\label{twomethods}Two calculation methods}

In this section, we outline the approaches used in the following to solve the given system. We start by characterizing the two parameter regimes studied throughout the later analysis. Then, we  describe two different methods to obtain the steady state solution for our model system. In the first method, we assume weak input fields. This allows us to approximate the infinite series of Fock states of each photon mode by the lowest few photon number states. This truncation of the Hilbert space renders the system finite, and thus allows us to numerically calculate a steady state density matrix. We call this method the truncated Hilbert space (TH) method, and discuss it in Sec.~\ref{fock}. 
In the second model, we assume the bad cavity regime for the resonator, such that the loss and photon scattering channels of the resonator dominate the system dynamics. Then, the cavity modes can be adiabatically eliminated. This method will be termed adiabatic elimination (AE) method, and is analyzed in Sec.~\ref{gencouplings}.
Finally, in Sec.~\ref{procontra} we compare the two methods and discuss their respective validity ranges.

\subsubsection{\label{parreg}Parameter regimes of interest}
We consider two fundamentally different parameter regimes~\cite{SargentMeystre}. On the one hand, we study the strong coupling case, in which the coupling between the atom and the cavity dominates the dynamics compared to the loss dynamics of the cavity modes. Thus, the conditions $\kappa_i\ll g_{m_i}$ are fulfilled. For the resonator-fiber part of the system, we assume critical coupling characterized by 
\begin{align}
\kappa_{i,ex}=\sqrt{h_i^2+\kappa_{i,in}^2}\,.
\end{align}
As the rates $\kappa_i$ contain the coherent scattering rates $h_i$ via $\kappa_{i,ex}$ as well as the incoherent photon decay rates $\kappa_{i,in}$, in the strong coupling regime, we also have $h_i< \kappa_i \ll g_{m_i}$. 
On the other hand, we consider the bad cavity regime. Here, the incoherent dynamics of the cavity modes dominates the dynamics, and the condition $g_{m_i}\ll \kappa_i$ holds. As before, then also $g_{m_i}\ll h_i$ is fulfilled.

\subsubsection{\label{fock}TH method - Fock mode truncation}

The general state of our system is determined by the state of the atom, as well as the state of the four different cavity modes. The dynamics can thus be described using basis states of the form
\be 
\ket{atom,\num(\mc  A_1),\num(\mc B_1),\num(\mc A_2),\num(\mc B_2)}\:,
\ee
where $\num(\mc M)$ 
denotes the number of photons in mode $\mc M$. If the parameter are chosen such that the mean occupation number of a cavity mode remains low, then it is possible to restrict the corresponding state space to low Fock number states.  Taking into account states with at most $l$ photons in each mode would lead to a state space reduced to dimension $3\times (l+1)^4$, since also the state without photons has to be considered for each mode. In the following, we assume the case in which at most one photon can be in each of the cavity modes. The number of basis states then reduces to $3\times 2^4 = 48$, and the corresponding reduced density matrix $\rho_{Fock}$ has $48^2=2304$ elements. Note that taking into account up to two excitations per mode would already lead to $59049$ density matrix elements, which is impractically large even though the number can be considerably reduced due to symmetries of the density matrix.

We arrange all elements of $\rho_{Fock}$ in a vector  and eliminate the last element using the trace condition $Tr(\rho_{Fock})=1$. The resulting vector $\vec\rho_{Fock}$ has $48^2-1$ elements, and we can rewrite the equations of motion for the density matrix elements as
\be 
\del{t}\vec\rho_{Fock}=\mc G\cdot \vec\rho_{Fock}+\vec K\,.
\ee
The steady state $\vec\rho_{Fock,stst}$ can then be obtained by solving the system of linear equations to give
\be 
\vec\rho_{Fock,stst}=-\mc G^{-1}\cdot \vec K\,.
\ee

\subsubsection{\label{gencouplings}AE method - Adiabatic elimination of the cavity modes}

In this approach, we assume that the dissipative dynamics of the cavity dominates the system dynamics, such that the degrees of freedom of 
the cavity can be adiabatically eliminated from the system~\cite{wismil}. In particular, we assume the bad cavity condition $g_{m_i}\ll \kappa_i$ to be fulfilled. 
After the elimination, we obtain an effective master equation for the atomic degrees of freedom only. This significantly reduces the system dimension 
and thus simplifies the solution. 

We apply standard techniques to eliminate the resonator modes~\cite{wismil}. Details on the calculation can be found in Appendix~\ref{adiab-details}. 
After the basis transformation Eqs.~(\ref{normalmodes}), we can treat the two normal mode pairs $\mc A_i$ and $\mc B_i$ separately. For each mode pair, 
first, we assume low occupation of the resonator modes due to the large dissipative dynamics, and expand the total density matrix in the photon 
occupation number. In the resulting density matrix equations, we assume 
\be 
\frac{|\mc L_{atom}|}{\kappa_i}\ll 1\,,
\ee
where $|\mc L_{atom}|$ represents the magnitude of terms corresponding to the coherent dynamics of the system, with $|\cdot|$ a suitable norm. Setting furthermore the time derivatives of off-diagonal elements in the expanded density matrix to zero, and inserting the thereby obtained expressions for the off-diagonal elements into the diagonal elements, we finally obtain an effective master equation for the atomic dynamics alone given by 
\begin{subequations}
\label{meqelim}
\begin{align}
\drhoA & =-\frac{i}{\hbar}[\HA,\rhoA] + \mc L_{\Gamma}\rhoA \,,\\
\mc L_{\Gamma}\rho &=  \sum_{i,j=1}^2\frac{\Gamma_{ij}}{2}\left(2\Sm{j}\rho\Sp{i}  -\Sp{i}\Sm{j}\rho \right. \nonumber \\
&\left. -\rho\Sp{i}\Sm{j}\right)\,, \\
\HA=&\hbar \sum_{i,j=1}^2\Delta_{ij}\Sp{i}\Sm{j} -\hbar \sum_{i=1}^2 \Delta_i\Sm{i}\Sp{i} \nonumber \\
&+ \hbar \Omega_i\Sp{i}+ \hbar \Omega_{i}^*\Sm{i}\:.
\label{elimham}
\end{align}
\end{subequations}
The constants $\Delta_{ij}$ and $\Omega_i$ are effective detunings and Rabi frequencies arising from the cavity elimination, respectively. Their values depend on the various coupling constants governing our system's dynamics. Explicit expressions can be found in Appendix~\ref{explcoupl}. Note that related results were obtained for a two-level atom coupled to a resonator with only one mode pair in~\cite{turnstile,turnstile}.

\subsubsection{\label{procontra}Comparison of the two calculation methods}

The adiabatic elimination method leads to an effective master equation, which can easily be solved and thus considerably reduces the computational effort. This comes at the price that the method only applies to the bad cavity limit, in which the dissipative dynamics dominates the system.
In contrast, the Fock state truncation method does not a priori exclude either the bad cavity or the strong coupling regime, but 
rather puts a constraint on the mean occupation number of the cavity modes alone. While potential resonant enhancement of the cavity 
occupation for low cavity loss rates must be taken into account, a weak input field generally suffices to remain within the validity range, 
such that this method also allows to access the strong coupling regime. Drawbacks of this method are the higher computational effort compared 
to the AE method, and the fact that by construction higher order correlation functions involving two photons are not accessible from the results. 
We further analyzed this point for a two-level atom coupled to the resonator, for which higher Fock modes can be included while keeping the total 
number of equations reasonably low. We found that it is not sufficient to take Fock states with two photonic excitations in total into account in order to reliably calculate the second order correlation function. Rather, one has to include two excitations per mode. As already mentioned, in the three-level atom case, including up to two photons per mode would increase the state space substantially, effectively rendering the calculations unfeasible.

Thus in summary, in the bad cavity regime, both methods can be used, and we verified that the two methods give the same results. Due 
to the reduced computational effort, here, the AE method will be used. In contrast, in the strong coupling regime, only the TH method can be applied.

\subsection{Input-Output Relations}
In this section, we relate the fields inside the resonator to the fields in the fiber. First, we derive an expression for the driving terms proportional to $\mc E_i$ in the equations of motion for the modes~(\ref{modes}). For this, we take the expectation value of Eqs.~(\ref{modes}), and obtain
\begin{align}
\label{adot1}
\mean{\del{t}{\mode{A}_i}}&= \mean{\del{t}\mc A_i}_{sys}- \frac{i\mc E_i}{\sqrt{2}}\:,
\end{align}
and equivalently for modes $\mc B_i$. The index $sys$ indicates all internal terms of the atom-cavity system not related to the external driving. 
Note that we  neglect the fluctuations of the input field in the driving terms entering the equations of motion for the modes by equating 
$\langle \mc E_i\rangle = \mc E_i$. On the other hand, written in terms of an external input field operator $\mode{A}_{i,in}$, the corresponding equation for $\mean{\del{t}\mode{A}_i}$ can be expressed as
\be
\mean{\del{t}{\mode{A}_i}}=\mean{\del{t}\mc A_i}_{sys} +\sqrt{\kappa_{i,ex}}\cdot \mean{\mode{A}_{i,in}}\:.\label{adot2}
\ee
From Eq.~(\ref{adot1}) and (\ref{adot2}), we can thus conclude 
\be 
\langle \mc A_{i,in}\rangle=-\frac{i\mc E_i}{\sqrt{2 \kappa_{i,ex}}}\:.
\ee
Note that if no input field is applied to the modes $b_i$, then $\langle\mc A_{i,in}\rangle=\frac{\langle a_{i,in}\rangle}{\sqrt{2}}$. 

Next, we consider the output from the resonator to the fiber. Using the input-output formalism~\cite{Gardiner1985,MilWals}, we find
\begin{subequations}
\begin{align}
a_{i,out}=&-a_{i,in}(t)+\sqrt{2\kappa_{i,ex}}a_i(t)\nonumber\\
=& - \langle a_{i,in}\rangle -a'_{i,in}(t) \nonumber \\
&+\sqrt{\kappa_{i,ex}} \left[ \mc A_i(t)+\mc B_i(t) \right]\,,\\
b_{i,out}=&-b_{i,in}(t)+\sqrt{2\kappa_{i,ex}}b_i(t)\nonumber\\
=&-b'_{i,in}(t)+\sqrt{\kappa_{i,ex}} \left[\mc A_i(t)-\mc B_i(t)\right]\:,
\end{align}
\end{subequations}
where $a'_{i,in}(t)$ and $b'_{i,in}(t)$ are fluctuations, such that for example $a_{i,in}=\langle a_{i,in}\rangle+a'_{i,in}(t)$.
As long as we consider normal ordered output fields we can substitute the output operators according to~\cite{turnstile}
\begin{subequations}
\label{sub}
\begin{align}
a_{i,out}&=\alpha_{i0}+\alpha_{i1}\Sm{1}+\alpha_{i2}\Sm{2}\,,\\
b_{i,out}&=\beta_{i0}+\beta_{i1}\Sm{1}+\beta_{i2}\Sm{2}\,.
\end{align}
\end{subequations}
Here, the complex coefficients $\alpha_{i0}$ and $\alpha_{i1}$ and $\alpha_{i2}$ are calculated as follows. We solve the equations of motion Eq.~(\ref{ABeom})
for the normal cavity modes $\mc A_i$ and $\mc B_i$ in the steady state. For the modes $\mc A_i$ the 
terms without an atomic operator we call $\alpha_{i0}$, the coefficient of $\Sm{1}$ in the equation 
for $\mc A_i$ we call $\alpha_{i1}$ and the terms in front of operator   $\Sm{2}$ are $\alpha_{i2}$. Analogously for the modes $\mc B_i$ we define 
the coefficients
as $\beta_{i0}$, $\beta_{i1}$ and $\beta_{i2}$.

\section{\label{observables}Observables}

\subsection{\label{fo}Transmission and reflection spectra}
The transmission spectra can be calculated from the normalized first order correlation functions as~\cite{glauber}
\be
\foc{m_i}=\frac{\mean{m_{i,\,out}^\dagger m_{i,\,out}}}{\mean{a_{i,\,out}^\dagger a_{i,\,out}}_{\Delta_i \gg \kappa}}
=\frac{\mean{m_{i,\,out}^\dagger m_{i,\,out}}}{\mean {a_{i,\,in}}^2}
\ee
for modes characterized by $m\in\{a,\,b\}$ and $i\in\{1,\,2\}$. The output photon operators can be related via Eqs.~(\ref{sub}) 
to the solution of equations describing the cavity-atom system. $\foc{a_i}$ can be interpreted as the transmission in dependence of the input 
field frequencies, whereas $\foc{b_i}$ is the corresponding reflection. The transmission and reflection can be calculated by both the AE and the 
TH methods introduced in Section~\ref{twomethods}, and are thus accessible in both the strong coupling and the bad cavity regime. 

\subsection{\label{so}Photon statistics}

In order to investigate the photon statistics of the output photons we have to consider the normalized second order correlation functions 
at equal times. They can be calculated as~\cite{glauber}
\be
\so{m_i,m_j}=\frac{\mean{m_{i,\,out}^\dagger m_{j,out}^\dagger m_{j,\,out} m_{i,\,out} }}{\mean{m_{i,\,out}^\dagger m_{i,\,out}} \mean{m_{j,\,out}^\dagger m_{j,\,out}}}\:.
\label{secor}
\ee
If $i=j$, the function $\so{m_i,m_i}$ gives the photon statistics of a single mode. If the correlation function is one, then the photons 
obey Poissonian statistics. Higher or lower values indicate super- or sub-Poissonian statistics, respectively. In case $i\neq j$, the function  
$\so{m_i,m_j}$ is a cross correlation function for photons arriving  at the detectors at the same time but in different modes $m_i$ and $m_j$.
As it was found in~\cite{turnstile} for a two-level atom coupled to a resonator, and as we also find in our analysis, the obtained photon statistics in the critical coupling regime is predominantly determined by the atom because the resonator contribution is suppressed by destructive interference in the critical coupling case. This suggests that sub-Poissonian statistics can be associated with anti-bunching, whereas super-Poissonian statistics can be connected to bunching. The second order correlation functions $\so{m_i,m_j}$ can only have positive values $\ge 0$. The slope of $\so{m_i,m_j}$ at time delay $\tau=0$ determines whether the photons exhibit bunching or anti-bunching. Bunching is indicated by a negative slope of $\so{m_i,m_j}$ whereas anti-bunching occurs for positive slopes. For $\so{m_i,m_j}=0$ the slope can only be non-negative and therefore in case of  $\so{m_i,m_j}=0$  we can conclude that we have anti-bunching.

Since for these second order correlation functions two photon processes are crucial, we can calculate them only by using the 
AE method introduced in Sec.~\ref{gencouplings}, but not via the TH method including only up to one-photon Fock states. Thus we 
only present results in the photon statistics in the bad cavity regime.

\section{\label{results}Results}

In this Section, we present the results  obtained numerically for our observables. As the system has many degrees of freedom, we have to make a number of parameter choices. Throughout the analysis, we assume that the coupling between different pairs of modes is weak and set $q=p=0$.  Furthermore, we consider the natural case that the two modes belonging to one WGM pair have similar properties, which leads to a number of parameter choices. First, we set $\kappa_{\mc A_1}=\kappa_{\mc B_1}=\kappa_1$ and $\kappa_{\mc A_2}=\kappa_{\mc B_2}=\kappa_2$ where $\kappa_i =\kappa_{i,ex}+\kappa_{i,in}$ and $\kappa_{i,ex}=\sqrt{\kappa_{i,in}^2+h_i^2}$. Similarly, for our calculations we assume similar properties of the two pairs of modes. We assume the prefactors in the atom-cavity coupling strengths Eqs.~(\ref{gs}) to be equal, i.e. $g_0^{(1)}=g_0^{(2)}\equiv g_0$. Note, however, that the different position-dependence still in general leads to different couplings of mode $\mc A_i$ and $\mc B_i$ to the two atomic transitions. For the coupling between resonator and fiber we set $\kappa_{ex,1}=\kappa_{ex,2}\equiv \kappa_{ex}$, for the atomic decay rate $\gamma_1=\gamma_2\equiv \gamma$ and for the scattering inside the cavity $h_1=h_2\equiv h$. The internal decay rate of the cavity modes are set to $\kappa_{i,in}=\gamma$, from which follows $\kappa_{1}=\kappa_{2}\equiv \kappa$. Furthermore, if not noted otherwise, we choose $k_ix=\pi/2$ which results in $g_{\mc A_i}=0$ and 
$g_{\mc B_i}=g_0$. Also, we assume that both the cavity mode pairs are on resonance with the atomic transition they couple to, respectively. 
Then, for the detuning $\delta_i=\Delta_i$ holds. The energy separation between the two ground states is set to $\delta=0$.

\subsection{\label{strong}Strong coupling regime}

In this section, we show our results for the transmission and reflection in the strong coupling regime. For the calculation 
we use the TH method described in Sec.~\ref{twomethods}. As parameters we choose $g_0=100\gamma$ and $h=15\gamma$. First, 
we present our results for the output flux of the different modes. Afterwards, we explain them using a dressed state model of the coupled atom.

\subsubsection{Transmission and reflection}

\begin{figure}[t]
\centering
\includegraphics[width=8cm]{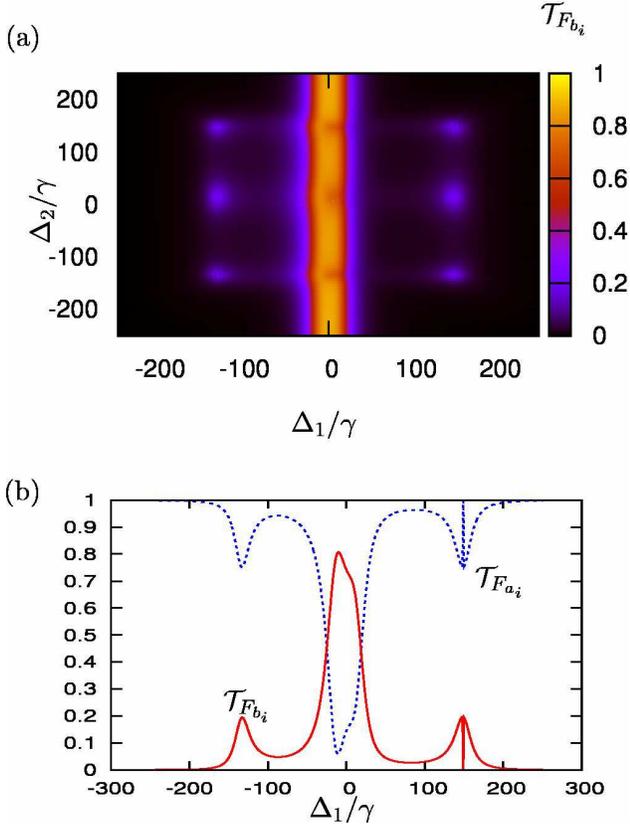}
\caption{\label{TFklein}(Color online) In (a) the normalized output flux $\foc{b_1}$ of mode $b_{1,out}$ in dependence of $\Delta_1$ and $\Delta_2$ is shown. (b) shows a cut of $\foc{a_1}$ and $\foc{b_1}$ at $\Delta_2=149\gamma$, which is 
 close to one of the eigenvalues of our system. The parameters are $(g_0,h,\kappa_{i,in})/\gamma=(100,15,1)$.
}
\end{figure}

\begin{figure}[t]
\centering
\includegraphics[width=8cm]{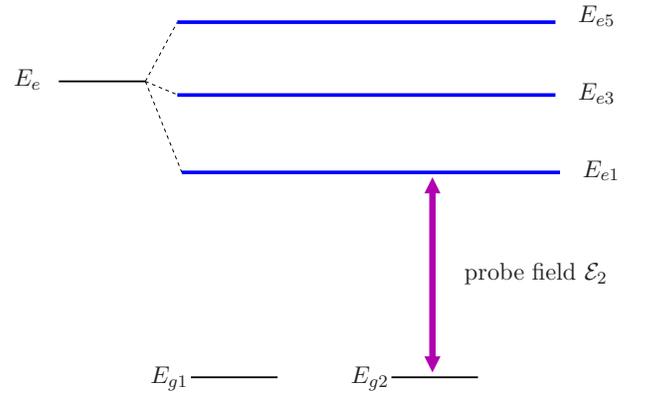}
\caption{\label{level}(Color online) Dressed state atomic level splitting for excitation with up to one cavity photon per mode. 
}
\end{figure}
\begin{figure}[t]
\centering
\includegraphics[width=8cm]{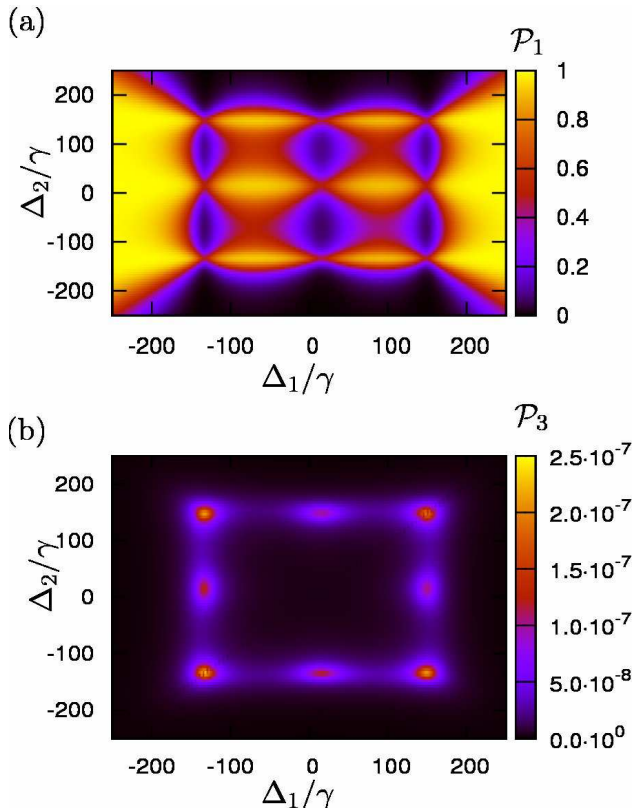}
\caption{\label{popklein}(Color online) In (a) the population of the atomic state 1 is shown in dependence of both the detunings. 
The parameters are $(g_0,h,\kappa_{i,in})/\gamma=(100,15,1)$. In (b) the population of state $\ket{3}$ is shown.}
\end{figure}

Fig.~\ref{TFklein} shows the transmission $\foc{a_1}$ and reflection $\foc{b_1}$ in dependence on the two probe field detunings $\Delta_1$ and $\Delta_2$ In the following, we analyze different regions in this figure and interpret the results with the help of the eigenstates of the system and the populations in the different atomic states. For this, the relevant state space determined below is shown in Fig.~\ref{level}, and the populations of the atomic states in Fig.~\ref{popklein}. 

If the absolute value of one of the detunings $|\Delta_i|$ is large, then the corresponding atomic transition $\ket{i}\rightarrow\ket{3}$ essentially decouples from the cavity. This leads to optical pumping into the atomic state $|i\rangle$, such that the corresponding population approaches one. For example, at  $\Delta_1=-250\gamma$ and $|\Delta_2|<|\Delta_1|$ the atom is  more likely excited from state $|2\rangle$ to state $|3\rangle$ than it is  from $|1\rangle$ to $|3\rangle$. Therefore, in the longtime limit the atom moves into state $|1\rangle$.  Eventually, the atom can be neglected, and the two mode pairs decouple. As for large detuning the light passes the resonator unperturbed, the reflection $\foc{b_1}$ is negligible.

A special situation arises if $\Delta_1=\Delta_2$. Then, both transitions of the atom are driven in the same way, and the atom evolves into a dark state~\cite{ScullyZubairy} with both ground states equally populated, and no population in the upper level $\pop_{3}$, as shown in Fig.~\ref{popklein}. This suppresses the interaction of the atom with the resonator, and thus the results on the main diagonal line $\Delta_1 = \Delta_2$ of Fig.~\ref{TFklein} are equal to the case without atom.

The further structures of the atomic populations $\pop_i$ and the reflection $\foc{b_1}$ can be understood from an eigenstate analysis of the matrix governing the equations of motion of our system. For this, we diagonalize the system's internal Hamiltonian  without external input field  and thus calculate the eigenenergies and the dressed eigenstates of the coupled cavity-atom system. We note that our parameter choice $\kappa_ix=\pi/2$ leads to $g_{\mc A_i}=0$ and thus only the modes $\mc B_i$ can couple to the atom. Then, the relevant basis states reduce to $|atom, \num({\mc B_1}),\num({\mc B_2})\rangle$. Including up to one photon per cavity mode,  five different eigenvalues and eleven eigenstates are obtained. The eigenvalues for the parameters used in Figs.~\ref{TFklein} and \ref{popklein} are 
\begin{subequations}
\label{EW}
\begin{align}
E_{g1}&=E_{g2}=0\,,\\
E_{e1}&=\frac{1}{2}\left(-h-\sqrt{8g_{\mc B_i}^2+h^2}\right)=-149\gamma\,,\\
E_{e2}&=-h=-15\gamma\,,\\
E_{e3}&=\frac{1}{2}\left(-h+\sqrt{8g_{\mc B_i}^2+h^2}\right)=134\gamma\,,\\
E_{e4}&=\frac{1}{2}\left(-3h+\sqrt{4g_{\mc B_i}^2+h^2}\right)=77\gamma\,,\\
E_{e5}&=\frac{1}{2}\left(-3h-\sqrt{4g_{\mc B_i}^2+h^2}\right)=-122\gamma\,,
\end{align} 
\end{subequations}
and the corresponding eigenstates read
\begin{subequations}
\begin{align}
\ket{V_{g,1}}&=|1,0,0\rangle\,, \qquad
\ket{V_{g,2}}=|2,0,0\rangle\,,\\
\ket{V_{e1}}&=|1,1,0\rangle+|2,0,1\rangle-\frac{iE_{e5}}{g_{\mc B_i}}|3,0,0\rangle\,,\\
\ket{V_{e2,1}}&=|2,1,0\rangle\,,\qquad
\ket{V_{e2,2}}=|1,0,1\rangle\,,\\
\ket{V_{e2,3}}&=-|1,1,0\rangle+|2,0,1\rangle\nonumber\\
\ket{V_{e3}}&=|1,1,0\rangle+|2,0,1\rangle-\frac{E_{e1}}{g_{\mc B_i}}|3,0,0\rangle\,,\\
\ket{V_{e4,1}}&=\frac{i(E_{e2}+2h)}{g_{\mc B_i}}|1,1,1\rangle+|3,0,1\rangle\,,\\
\ket{V_{e4,2}}&=\frac{i(E_{e2}+2h)}{g_{\mc B_i}}|2,1,1\rangle+|3,1,0\rangle\,,\\
\ket{V_{e5,1}}&=\frac{i(E_{e4}+2h)}{g_{\mc B_i}}|1,1,1\rangle+|3,0,1\rangle\rangle\,,\\
\ket{V_{e5,2}}&=\frac{i(E_{e4}+2h)}{g_{\mc B_i}}|2,1,1\rangle+|3,1,0\rangle\rangle\,,
\end{align}
\end{subequations}
We can identify three different kinds of eigenstates. The first group contains the ground states $\ket{V_{g,1}}$ and $\ket{V_{g,2}}$, 
with the atom in one of its two ground states and no photon in the cavity. Since we typically apply weak probe fields, our coupled system 
is mostly in one of these two states. 
The second group contains all states containing a single excitation of the system. The eigenstates belonging to this group are $\ket{V_{e1}}$, $\ket{V_{e2,1-3}}$ and $\ket{V_{e3}}$. These three states $\ket{V_{e2}}$ are the states in which the atom is in one of the ground states but the cavity is populated by one photon. The states $\ket{V_{e1}}$ and $\ket{V_{e3}}$ are superpositions including also the state with excited atom and no photon in the cavity modes.
The third group contains states in which the system contains two excitations, one of mode $\mc B_1$ and one of mode $\mc B_2$. It should be noted that these states cover only a small part of the doubly excited eigenstate space due to our initial choice of considering at most one photon per mode. But due to the low excitation considered, the doubly excited states have only negligible weak influence on the system dynamics and therefore can be neglected for the interpretation. The states including up to one excitation are shown in Fig.~\ref{level}. 

To interpret the system, it is important to note that due to our definition of the detunings, the probe fields are resonant to an atomic dressed state transition if the detuning has the opposite sign of the respective eigenenergy shift.
First, we focus on the line defined by $\Delta_2=149\gamma$, for which the probe field $\mc E_2$ is resonant with the transition 
$\ket{V_{g,2}}\rightarrow \ket{V_{e1}}$. Fig.~\ref{TFklein}(b) shows the transmission $\foc{a_1}$ and the reflection $\foc{b_1}$ at this detuning.  
For $\Delta_1=-250\gamma$, the condition $|\Delta_2|\ll|\Delta_1|$ holds and thus $\foc{b_1}\approx 0$ and the atom is almost completely 
in state $\ket{1}$, see Fig.~\ref{popklein}. 
Scanning $\Delta_1$ towards zero detuning, a resonance structure with increased reflection can be seen at around $\Delta_1=-134\gamma$. At 
this point, the field $\mc E_1$ is near-resonant with the transition $\ket{V_{g,1}}\rightarrow \ket{V_{e3}}$, which leads to a scattering of 
photons in the reflection channel via the atom. This can also be seen by the increase in the excited state population of the atom, see 
Fig.~\ref{popklein}. 
Further increasing $\Delta_1$, the reflection becomes minimal at $\Delta_1\approx -80\gamma$. In this region, field $\mc E_1$ is relatively far detuned from the eigenstates of the system, as can be seen from the high population of the atomic state $|1\rangle$. As $\pop_3$ is relatively small, also the reflection remains low. 
At $\Delta_1=-15\gamma=-h$, a strong resonance is observe in the reflection spectrum. This peak can be attributed to the non-coupling cavity mode $\mc A_1$. A closer analysis shows that the structure around $\Delta_1=-15\gamma=-h$ in fact contains two separate peaks, which leads to the apparent asymmetry. The second peak is located at $\Delta_1=15\gamma=h = -E_{e2}$ and is due to the coupling to the eigenstates $|V_{e2}\rangle$. 
At positive detunings $\Delta_1$,  we first encounter a region with low reflection, which can be interpreted in the same way as the region around $\Delta_1 = -80\gamma$. 
At detunings around $\Delta_1=149\gamma$, both probe fields couple to the eigenstate  on the transition $\ket{V_{g,1}}\rightarrow \ket{V_{e1}}$. At exact two-photon resonance $\Delta_1=\Delta_2$, a narrow dip in the reflection can be seen, which is due to the above-mentioned dark state. In this case the system behaves as if the atom was not present. In the vicinity of the dark state, the population of the excited state is relatively high due to the near-resonant coupling to the eigenstate $|V_{e1}\rangle$, which leads to the increase in reflection. 

Next, we analyze the region around $\Delta_1\approx 0$ in Fig.~\ref{TFklein}(a), where high reflectivities occur. It can be seen that this structure has a small residual dependence on $\Delta_2$. At detunings $\Delta_2$ for which the second field is in resonance with one of the system's eigenstates, the reflection resonance is centered around $\Delta_1 =-h= - 15 \gamma$. For all other values of $\Delta_2$, the structure is centered approximately around $\Delta_1 = 0$.
The reason for this is as follows. If the second field is detuned from all resonances, then the atom is optically pumped into state $|V_{g2}\rangle$ by the first field, and the system again acts as if no atom was present.  Without atom, the two cavity mode pairs are symmetrically split due to their coupling $h$ such that they have resonances at $\pm h = \pm 15\gamma$. As these two resonances are close to each other compared to the line width, they overlap and appear as a single resonance centered around $\Delta_1 = 0$. 
In contrast, if $\Delta_2$ is tuned to resonance with one of the system's eigenstates, then the atom contributes to the system's dynamics. Then, in the reflection spectrum the non-coupling mode at $\Delta_2 = -15\gamma$ dominates, as already explained for the central resonance of Fig.~\ref{TFklein}(b).  Considering the atomic population of the excites state $\pop_3$ in Fig.~\ref{popklein} we observe a peak for $\Delta_1=0$ at $\Delta_2=134\gamma$ and $\Delta_2=-149\gamma$ whereas for the third eigenvalue $\Delta_2=15\gamma$ the population $\pop_3\approx 0$. This is because the eigenstates $V_{e2,1-3}$ do not include any state in which the atom is excited.

\subsubsection{\label{applstrong}Dependence on the atom position}

In this Section, we analyze the dependence of the results on the position of the atom. In general, the position of the atom determines the coupling constants $g_{\mc A_i}$ and $g_{\mc B_i}$, which in turn influence the spectral properties of the total system. A potential realization for the following analysis could be an atom trapped in a potential created by the evanescent field of the resonator. This would lead to a trapping in radial direction, but could still allow for movements of the particle along the azimuthal direction. 

\begin{figure}[t]
\centering
\includegraphics[width=8cm]{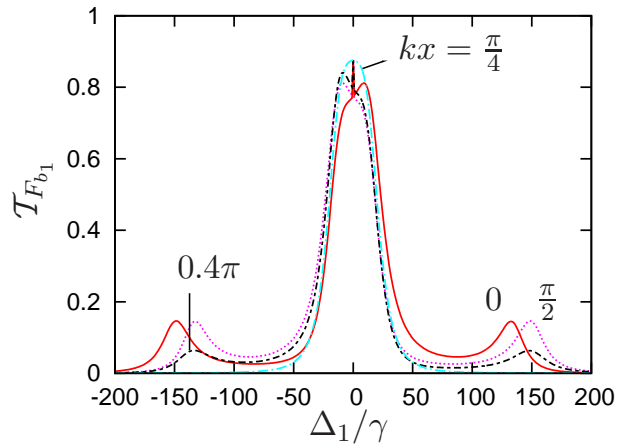}
\caption{\label{position}(Color online) Reflection $\foc{b_1}$ for different azimutal positions $x$ of the atom and fixed $\Delta_2=0$. We chose $kx=0$, $\frac{\pi}{4}$, $0.4 \pi$, and $\frac{\pi}{2}$.}
\end{figure}

In Fig.~\ref{position}, we show our results for $\Delta_2=0$ for different positions of the $\Lambda$ atom. Here, the distance to the cavity is constant but the azimuthal position along the resonator circumference and thus the phase $k_i x$ is varied. If $k_i x=\pi/2$, the atom couples to modes $\mc B_i$ only as assumed for the previously shown results. In case of $k_i x=0$, the atom couples to the $\mc A_i$ modes only. Intermediate values lead to coupling to both modes. Changing $k_i x$ from $0$ to $\pi/2$ corresponds to moving the atom over a distance of $\lambda/4$ along the circumference. It can be seen in Fig.~\ref{position} that while the general structure of the spectrum remains constant, the positions of both sidebands oscillate around their mean position. Also the height of the sideband peaks changes, with maxima for the two cases in which a single mode pair couples to the atom. At $k_i x=\pi/4$, the sideband peaks vanish.
Additionally, the slight asymmetry in the central resonance around $\Delta_1=0$ changes. For example, in moving from $k_i x=0$ to $k_i x= \pi/2$, the structure becomes mirrored at the $\Delta_1=0$ axis. This is because the modes $\mc A_i$ and $\mc B_i$  exchange their roles as coupled and uncoupled modes. For $k_i x=0.4\pi$, both mode pairs $\mc A_i$ and $\mc B_i$ couple to the atom, but with different coupling strengths. For $k_i x=\pi/4$ the atom couples with equal strength to the $\mc A_i$ and $\mc B_i$ modes. Therefore, the middle peak structure is symmetric around $\Delta_1=0$. 

The radial dependence is directly governed by the absolute value of the coupling constants $g_{\mc A_i}$ and $g_{\mc B_i}$. The peaks in the spectrum lie at the eigenenergies given in Eq.~(\ref{EW}). For fixed $\Delta_2$ they move to higher values of $|\Delta_1|$ when increasing the strength of the coupling between the atom and the cavity.

This dependence of the spectral properties on the particle position can also be used to detect the presence of the particle, or to measure its position.
 
\subsection{\label{badcav}Bad cavity regime}

In this Section we analyze the bad cavity regime, and choose $g_0=70\gamma$ and $h=250\gamma$, and all other parameters are chosen as in Sec.~\ref{strong}. Our observables are the output fluxes and the second order correlation functions. For the calculations we use the AE method as described in Sec.~\ref{gencouplings}.

\subsubsection{Transmission and reflection}

\begin{figure}[t]
\centering
\includegraphics[width=8cm]{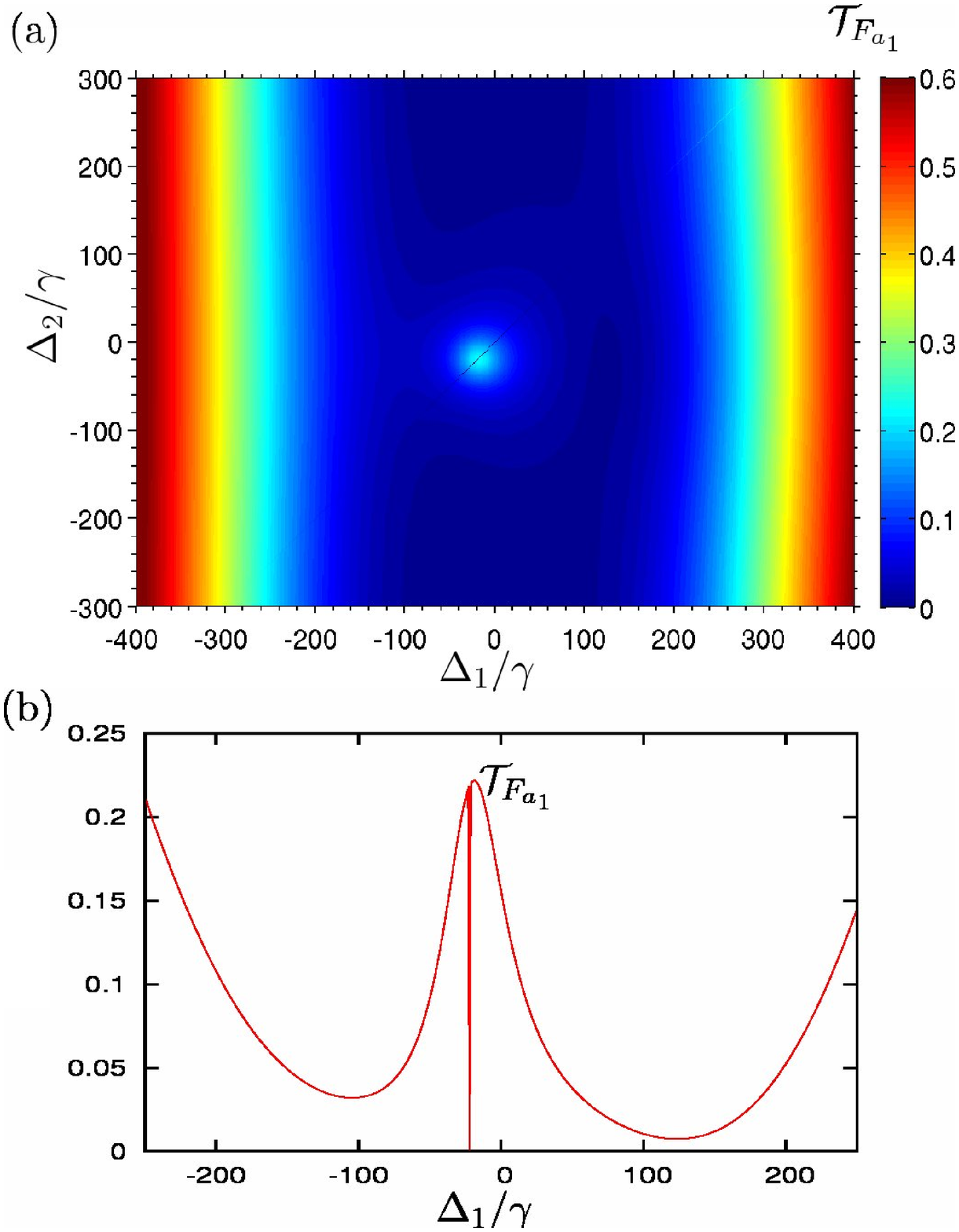}
\caption{\label{TFgross}(Color online) (a) $\foc{a_1}$ for the bad cavity regime with $g_0=70\gamma$ and $h=250\gamma$. (b) Cut at $\Delta_2=-22\gamma$ for the same parameters.}
\end{figure}

We start by analyzing the transmission into mode $a_{1,out}$ shown in Fig.~\ref{TFgross}. As in the strong coupling case, 
the transmission approaches one if the first probe field is far detuned, i.e., for large $|\Delta_1|$. Residual interactions 
lead to a trapping of the atom in state $|1\rangle$ if the second field is less detuned. Decreasing the absolute value of $\Delta_1$, 
the transmission reduces and approaches zero almost independently from the second detuning $\Delta_2$. 
There is a single resonance structure around $\Delta_1 = -19\gamma$ and $\Delta_2 = -19\gamma$ with increased transmission $\foc{a_1}$. A cut through Fig.~\ref{TFgross}(a) at $\Delta_2=-22\gamma$ is shown in Fig.~\ref{TFgross}(b). A closer analysis revealed that both the position and the amplitude of this peak depend on the atom position via the phase $k_i x$. In Fig.~\ref{TFgross}, $k_i x=\pi/2$ which means that modes $\mc B_i$ couple to the atom whereas modes $\mc A_i$ do not couple. On the other hand, choosing $k_i x=0$ such that $\mc A_i$ couples and $\mc B_i$ not, the resonance structure moves to $\Delta_1 = 19\gamma$, $\Delta_2 = 19\gamma$, and has the same intensity as for $k_i x=\pi/2$. For values of $k_i x$ between $0$ and $\pi/2$ the resonance moves along the diagonal line and has reduced amplitude. For equal coupling of the two modes, $g_{\mc A}=g_{\mc B}$, the resonance vanishes. We can thus directly trace the resonance back to the coupled atom.  A similar resonance was also found for a two level atom coupled to a resonator in the strong coupling regime~\cite{turnstile}.
Interestingly, however, the resonance is not located at one of the eigenstates of the system obtained by diagonalizing the coherent part of the initial interaction Hamiltonian. Rather, the resonance coincides with maximum population in the atom's upper state. This can be explained by diagonalizing the effective Hamiltonian obtained after the adiabatic elimination in Eq.~\ref{elimham}. The only nonvanishing eigenvalue of the system after the adiabatic elimination is equal to the sum of the corrected effective detunings $\Delta_{11}+\Delta_{22}$. For our parameters $\Delta_{11}=\Delta_{22}=\frac{g_{\mc B}^2h} {h^2+\kappa^2}$. Inserting the parameters yields an expected position of the resonance of $\Delta_1=\Delta_2=-19.5\gamma$.
Finally, on the diagonal line $\Delta_1=\Delta_2$, the atom is in a dark state such that it decouples from the resonator. Thus the 
transmission is as in the case without atom for these parameter values. In the lower subfigure, the dark state leads to the sharp dip 
around $\Delta_1=-22\gamma$.

 
\subsubsection{Second order correlation functions}

\begin{figure}[t]
\centering
\includegraphics[width=8cm]{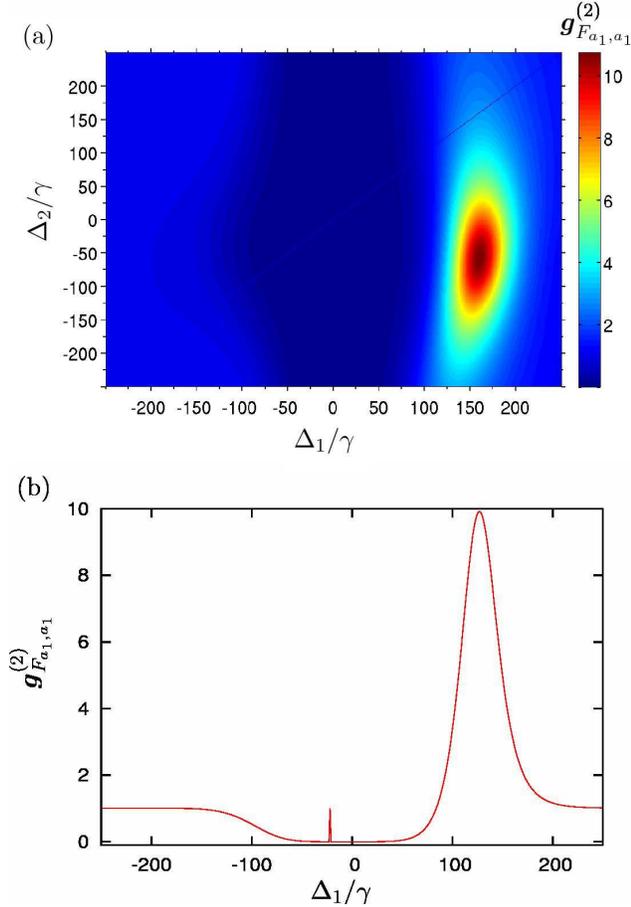}
\caption{\label{gFgross}(Color online) (a) Second order correlation function $\so{a_1,a_1}$ for the bad cavity regime with $g_0=70\gamma$ and $h=250\gamma$. (b) Cut at $\Delta_2=-22\gamma$ for the same parameters.}
\end{figure}

In this Section, we analyze the second order correlation functions $\so{m_i,m_j}$. First, we consider the case $i=j$. In Fig.~\ref{gFgross}, we show the second order correlation function for mode $a_{1,out}$. In the lower subfigure, a cut at $\Delta_2=-22\gamma$ is shown. 
The dominant feature is a peak around $\Delta_1 = 130\gamma$ and $\Delta_2 = -30 \gamma$. Around this position,  $\so{a_1,a_1}$ reaches high values up to 10. Thus, for these detunings the photons obey a super-Poissonian statistics. This structure is due to  small values of $\foc{a_i}$ for these detunings. Since the correlation function $\so{a_1,a_1}$ is normalized to the output flux, see Eq.~\ref{secor}, small intensities lead to an enhancement of  $\so{a_1,a_1}$.  
Another important feature is the large area around zero detunings in which the system exhibits sub-Poissonian statistics. This area coincides with the resonance in the transmission around $\Delta_1 = -19\gamma$, $\Delta_2=-19\gamma$ found in Fig.~\ref{TFgross}. In this parameter regime, non-classical light with sub-Poissonian photon statistics is generated. Since this second order correlation function has values around zero, we can conclude from this information that we have anti-bunching. The non-classical light arises from the interaction with the atom, similar to the turnstile operation reported in~\cite{turnstile}. Without atom, the transmission is negligible due to destructive interference in forward direction downstream of the resonator. But excess excitations emitted from the atom can be transmitted in forward directions, and these exhibit sub-Poissonian statistics and anti-bunching due to the fact that the atom requires a finite time to be excited.
Finally, in the dark state case $\Delta_1=\Delta_2$, the photon statistics is Poissonian, $\so{a_1,a_1}=1$.

\begin{figure}[t]
\centering
\includegraphics[width=8cm]{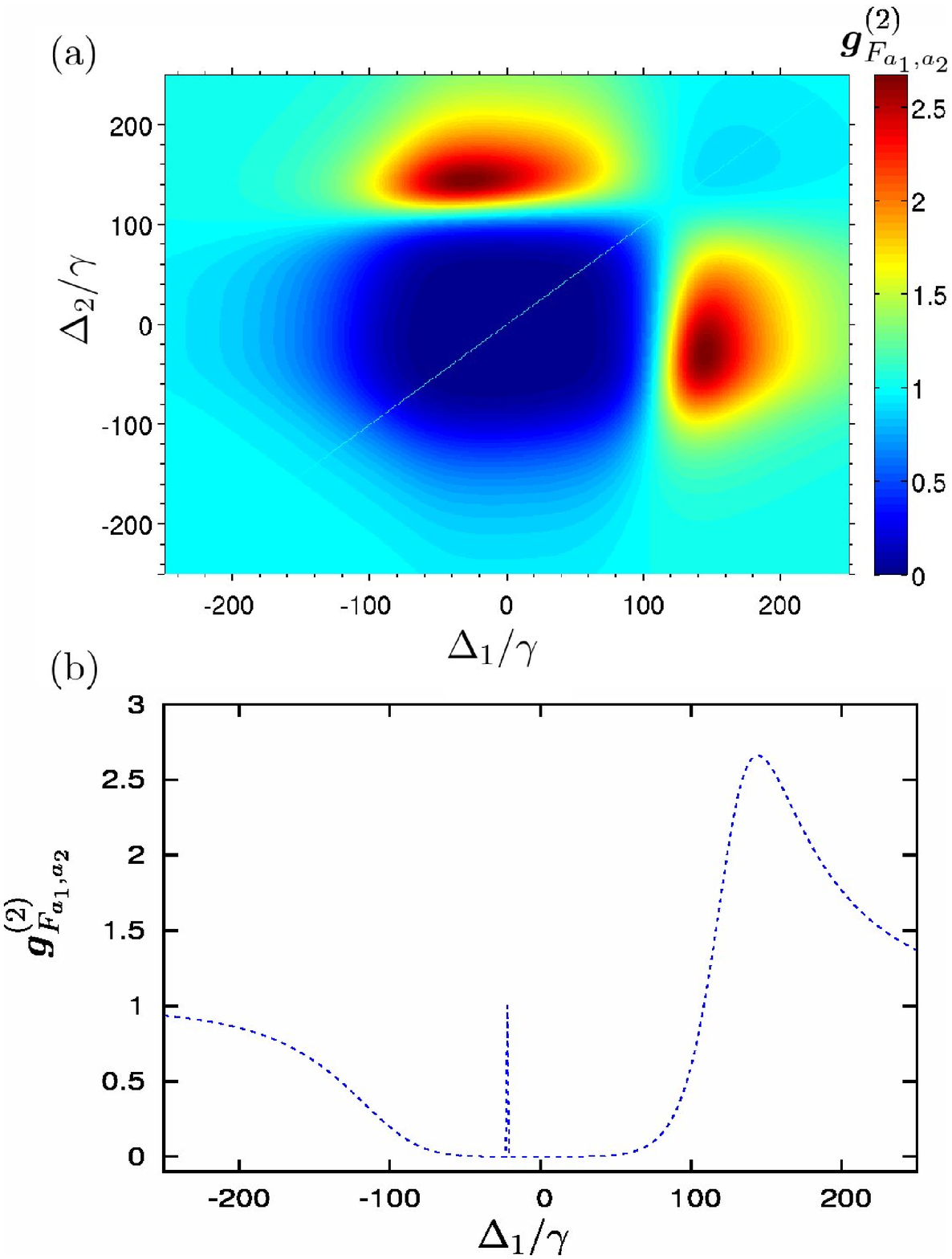}
\caption{\label{crgross}(Color online) (a) Cross correlation $\so{a_1,a_2}$ for the bad cavity regime with $g_0=70\gamma$ and $h=250\gamma$. (b) Cut at $\Delta_2=-22\gamma$ for the same parameters.}
\end{figure}

Next, we considered the cross correlation functions. In Fig.~\ref{crgross}, the cross correlation of mode $a_{1,out}$ and $a_{2,out}$ is presented. In can be seen that it also is close to zero in the region around  $\Delta_1 = -19\gamma$, $\Delta_2=-19\gamma$  which is the position of the transmission resonance  in $\foc{a_1}$. This extends the turnstile operation to two distinct modes $a_{1,out}$ and $a_{2,out}$. Thus a detector placed in the transmission direction only detects individual photons in either of the two modes. We also observe two regions with high second order correlation $\so{a_1,a_2}$ in Fig.~\ref{crgross}(a). Analogously as for $\so{a_1,a_1}$, this can be explained by minima in either of the two output intensities to which the correlation function is normalized. In the region around $(\Delta_1,\Delta_2)=(130,-30)\gamma$ the first order correlation $\foc{a_1}\approx 0$. In the second peak region around $(\Delta_1,\Delta_2)=(-30,130)\gamma$ the transmission of mode $a_2$ almost vanishes. 
%
 
\subsubsection{\label{apply}Applications}

In this Section, we outline possible applications of our setup in the bad cavity regime. First, we show that the coupled 
cavity atom system can be used as a single photon source. Next, we show that it acts as a bi-modal photon turnstile. Finally, 
we show that our bimodal photon turnstile is tunable by the input fields and their detunings. This could be exploited as an optical switch.

Considering Fig.~\ref{gFgross}(a), we see that in a region of the resonance in $\foc{a_1}$ close to 
$(\Delta_1,\Delta_2)=(-19,-19)\gamma$ (c.f. Fig.~\ref{TFgross}) the second order correlation $\so{a_1,a_1}$ vanishes. This means that photons 
of mode $a_{1,out}$ leave the system individually, such that it is impossible to  detect two photons of this mode at the same time. Moreover, 
the cross correlation $\so{a_1,a_2}$ also vanishes, see Fig.~\ref{crgross}(a). This means, the photons of each mode come out individually and 
additionally, if an $a_1$ photons is detected no $a_2$ photon can be detected at the same time. Consequently, whenever an $a_{1,out}$ photon 
is detected, it is impossible that an $a_{2,out}$ photons arrives at the detector at the same time. Therefore we know, that in $a$ direction only 
individual photons leave the systems. The output intensity of both transmission modes $a_{i,out}$ is different from zero $\foc{a_1}\neq 0$ and 
$\foc{a_2}\neq 0$, see  Fig.~\ref{TFgross}. Thus our coupled system can be used as bi-modal single photon source in $a_i$ direction. 

This can also be used as a photon turnstile, in $a$ direction. Photons come out individually and turn their propagation direction in case of 
non vanishing atom cavity interaction, which means that the atom is not in a kind of dark state. The turnstile can be controlled by changing 
the detunings and therewith driving the atom into a dark state. Controllable here means that we can turn the output flux on or off simply by 
changing the detuning of one of the input probe fields. For example, if we choose $\Delta_2=-19\gamma$ and $\Delta_1\approx-19\gamma$ but 
$\neq -19\gamma$. Then the transmission intensities $\foc{a_1}$ and $\foc{a_2}$ are almost maximum and thus $\neq 0$. However, when slightly 
tuning the input laser field $\mc E_1$ to $\Delta_1=\Delta_2=-19\gamma$, the atom is in the dark state and thus does not interact with the cavity any more. 
Then the system behaves as if no atom was there and the transmission of both the modes is almost zero. Thus, whether we have a nonvanishing 
transmission intensity or not can be controlled by slightly changing either of the input detunings.

Furthermore, an optical switch for the modes $a_{i,out}$ can be realized by our system in the bad cavity regime. If the parameters are such that the atom is effectively decoupled from the resonator,  due to the critical coupling condition no incoming $a_{i,in}$ photon can be transmitted. Thus all incoming photons are reflected back from the cavity. But enabling the atom-cavity coupling, some photons are transmitted and thus the flux in $a$ direction becomes non-zero. 
As an example, we found that for $\Delta_1\approx-19\gamma$ but $\Delta_1\neq -19\gamma$ and $\Delta_2=-22\gamma$ the transmission $\foc{a_1}$ is maximum, see Fig.~\ref{TFgross}(b). But if the input field $\mc E_2$ is switched off, the atomic population is completely transferred to ground state $\ket{2}$ such that the atom decouples from the resonator. Then, the transmission $\foc{a_1}\approx  0$, and all light is reflected.

\section{\label{summ}Summary and Discussion}

We presented a system consisting of a tapered glass fiber and a toroidal microcavity coupling to a 3-level atom in $\Lambda$ configuration. We considered two different parameter regimes, the strong coupling regime and the bad cavity regime. For both these regimes we presented adequate calculation methods in order to solve the equations of motion for the system's operators, namely, one based on a truncation of the Hilbert space, and one based on an adiabatic elimination of the resonator. We compared the respective validity ranges, and their advantages and disadvantages. Our observables are the transmission and reflection output fluxes as well as second order correlation functions and cross correlations and the atomic populations. We found that our system can work as a tunable bimodal photon turnstile which can be controlled by the detuning of the two input fields and explained how to exploit our setup as a photonic switch. Additionally, we analyze the dependence of our observables on the position of the nearby atom. In both the strong and weak coupling regime, we explained the structures of the transmission and reflection using respective dressed states or eigenmodes of the system. 

In our calculations, we neglected direct interactions between different mode pairs in the resonator, and assumed selective coupling of the two atomic transitions to one mode pair, respectively. This could be realized by using the similar frequency, but different polarizations for the two mode pairs which, however, is technically challenging. Alternatively, different frequencies and similar polarizations could be used, and the atom could be tuned in resonance with both modes by adjusting a Zeeman shift using an external magnetic field. Finally, our system can also be realized by two cavities coupling to the same $\Lambda$ atom.  This alternative implementation for the considered setup is shown in  Fig.~\ref{twocav}. In this setup, the atom couples to two resonators, and each mode pair is restricted to one of the resonators. As the modes in the two resonators have different polarization or different frequencies, direct coupling between the resonators could be avoided. This setup at the same time could  provide a connection to studies on quantum transport of electrons through single molecules~\cite{electronics}.

\begin{acknowledgments}
Financial support by the International Max Planck Research School for Quantum Dynamics in Physics, Chemistry and Biology is appreciated.
\end{acknowledgments}

\appendix

\section{\label{adiab-details}Details on the adiabatic elimination}
We start with the Hamiltonian given in Eq.~(\ref{ABham}) and the
complete master equation including all cavity modes and the atomic states
\be 
\del{t}\rho=-\frac{i}{\hbar}[\HAB,\rho]+\mc L_{\kappa}\rho+\mc L_{\gamma}\rho
\label{cme2}
\ee
with the Liouville superoperators
\be
\mc L_{\kappa}\rho=\sum_{\mc M\in \{\mc A_i, \mc B_i\}}\kappa_{\mc M}(2\mc M\rho\mc M^\dagger-\mc M^\dagger \mc M\rho-\rho\mc M^\dagger \mc M)
\ee
and
\be 
\mc L_{\gamma}\rho=\sum_{i=1}^2\frac{\gamma_i}{2}(2\Sm{i}\rho\Sp{i}-\Sp{i}\Sm{i}\rho-\rho\Sp{i}\Sm{i})\:.
\ee
Due to the weak input field and the fast decay of the cavity population, we can assume that the average occupation number is low. 
Furthermore, after the basis transformation Eq.~(\ref{normalmodes}), we can treat the two normal mode pairs $\mc A_i$ and $\mc B_i$ 
separately. We show the elimination formalism with modes $\mc A_i$ in the following. We start by expanding the density matrix in the photon number 
occupation, keeping terms including up to one photon per mode. The expanded density operator reads~\cite{wismil}
\begin{align}
\rhoexp=&\rho_{00}\otimes\ket{00}\bra{00}+\rho_{10}\otimes\ket{10}\bra{00}\nonumber\\
&+\rho_{20}\otimes\ket{01}\bra{00}+\rho_{01}\otimes\ket{00}\bra{10}\nonumber\\
&+\rho_{02}\otimes\ket{00}\bra{01}+\rho_{03}\otimes\ket{00}\bra{11}\nonumber\\
&+\rho_{11}\otimes\ket{10}\bra{10}+\rho_{12}\otimes\ket{10}\bra{01}\nonumber\\
&+\rho_{21}\otimes\ket{01}\bra{10}+\rho_{22}\otimes\ket{01}\bra{01}\nonumber\\
&+\rho_{03}\otimes\ket{00}\bra{11}+\rho_{30}\otimes\ket{11}\bra{00}\nonumber\\
&+\rho_{31}\otimes\ket{11}\bra{10}+\rho_{32}\otimes\ket{11}\bra{01}\nonumber\\
&+\rho_{33}\otimes\ket{11}\bra{11}+\rho_{23}\otimes\ket{01}\bra{11}\nonumber\\
&+\rho_{13}\otimes\ket{10}\bra{11}\:,
\label{rhoexp}
\end{align}
where the photon states for the two modes of the $\mc A_i$ pair are defined as $\ket{\num(\mc  A_1),\num(\mc A_2)}$. The $\rho_{ij}$ are the residual part of the density matrix. 

Substituting the expanded density matrix Eq.~(\ref{rhoexp}) into the master equation~\ref{cme2}, we obtain equations of motion for the elements $\rho_{ij}$. The equations corresponding to the lowest photon number occupation evaluate to 
\begin{subequations}
\label{rhodot}
\begin{align}
\del{t}\rho_{00}=&\mc L_{atom}\rho_{00}+2\kappa_1\rho_{11}+2\kappa_{2}\rho_{22}\nonumber\\
&-i\left(\left(\frac{1}{\sqrt{2}}\mc E^*_1+g_{\mc A_1}\Sp{1}\right)\rho_{10}\right.\nonumber\\
&+\left(\frac{1}{\sqrt{2}}\mc E^*_2+g_{\mc A_2}\Sp{2}\right)\rho_{20}\nonumber\\
&-\rho_{01}\left(\frac{1}{\sqrt{2}}\mc E_1+g_{\mc A_1}\Sm{1}\right)\nonumber\\
&\left.-\rho_{02}\left(\frac{1}{\sqrt{2}}\mc E_2+g_{\mc A_2}\Sm{2}\right)\right)\:,\\
\del{t}\rho_{10}=&\mc L_{atom}\rho_{00}-\kappa_{1}\rho_{10}\nonumber\\
&-i\Big(\left(\delta_1+h_1\right)\rho_{10}\Big.\nonumber\\
&+\left(\frac{1}{\sqrt{2}}\mc E_1+g_{\mc A_1}\Sm{1}\right)\rho_{00}\nonumber\\
&-\rho_{11}\left(\frac{1}{\sqrt{2}}\mc E_1+g_{\mc A_1}\Sm{1}\right)\nonumber\\
&-\rho_{12}\left(\frac{1}{\sqrt{2}}\mc E_2+g_{\mc A_2}\Sm{2}\right)\nonumber\\
&\left.+\left(g_{\mc A_2}\Sp{2}+\frac{1}{\sqrt{2}}\mc E_2^*\right)\rho_{30}+(m+p) e^{i\epsilon t}\rho_{20}
\right)\:,\\
\del{t}\rho_{11}=&\mc L_{atom}\rho_{11}-2\kappa_1\rho_{11}\nonumber\\
&-i\left(\left(\frac{1}{\sqrt{2}}\mc E_1+g_{\mc A_1}\Sm{1}\right)\rho_{01}\right.\nonumber\\
&\left.-\rho_{10}\left(\frac{1}{\sqrt{2}}\mc E^*_1+g_{\mc A_1}\Sp{1}\right)\right)\:.
\end{align}
\end{subequations}
Here, $\mc L_{atom}\rho$ is defined as
\begin{align}
\mc L_{atom}\rho &=-\frac{i}{\hbar}[\mc H_0,\rho]+\mc L_\gamma\rho\:,\\
\mc H_0&=- \hbar \sum_{i=1}^2 \Delta_i \Sm{i} \Sp{i}\:.
\end{align}

In the bad cavity regime, the cavity loss dynamics dominates over the coherent dynamics, such that
\be 
\frac{|\mc L_{atom}|}{\kappa_i}\ll 1\,,
\ee
and thus the terms $\mc L_{atom}$ can be neglected in Eqs.~(\ref{rhodot}). Setting further the time derivative of the diagonal elements in Eqs.~(\ref{rhodot}) to zero, we arrive at 
\begin{align}
\rho_{10}=&\frac{-i}{\kappa_1+i(\delta_1+h_1)}\nonumber\\
&\cdot\left((\frac{1}{\sqrt{2}}\mc E_1+g_{\mc A_1}\Sm{1})\rho_{00}\right.\nonumber\\
&-\rho_{11}(\frac{1}{\sqrt{2}}\mc E_1+g_{\mc A_1}\Sm{1})\nonumber\\
&-\rho_{12}(\frac{1}{\sqrt{2}}\mc E_2+g_{\mc A_2}\Sm{2})\nonumber\\
&\left.+(g_{\mc A_2}\Sp{2}+\frac{1}{\sqrt{2}}\mc E_2^*)\rho_{30}+(m+p) e^{i\epsilon t}\rho_{20}
\right) \:.
\end{align}

Now we calculate the time evolution of the atomic density matrix using the diagonal elements of $\del{t}\rhoexp$
\be
\del{t}\rhoA=\del{t}\rho_{00}+\del{t}\rho_{11}+\del{t}\rho_{22}\:.
\ee

Since our cavity is almost always empty, we substitute the terms obtained for the off diagonal elements $\rho_{ij}$ only up to $\rho_{00}$ order. 

Therewith we finally arrive at an atomic master equation of the form as shown in Eq.~(\ref{elimham}).
\\

\section{\label{explcoupl}Constants for the Atomic Master Equation}
The constants used in order to calculate the atomic master equation Eq.~(\ref{elimham}) are defined as follows:
\begin{align}
&\Delta_{11}&=&g_{\mc A_1}^2\Re(\lambda_{\mc A})+g_{\mc B_1}^2\Re(\lambda_{\mc B})\nonumber\\
&\Delta_{22}&=&g_{\mc A_2}^2\Re(\xi_{\mc A})+g_{\mc B_2}^2\Re(\xi_{\mc B})\nonumber\\
&\Delta_{12}&=&\Delta_{21}=0\nonumber\\
&\Omega_1&=&\Omega_{\mc A_1}+\Omega_{\mc B_1}\nonumber\\
&\Omega_2&=&\Omega_{\mc A_2}+\Omega_{\mc B_2}\nonumber\\
\ix{with}\nonumber\\
&\Omega_{\mc A_1}&=&\lambda_{\mc A}^*g_{\mc A_1}\frac{\mc E_1}{\sqrt 2}+\mc F_{\mc A}^*\mu g_{\mc A_1}\frac{\mc E_2}{\sqrt 2}\nonumber\\
&\Omega_{\mc A_2}&=&\xi_{\mc A}^*g_{\mc A_2}\frac{\mc E_2}{\sqrt 2}+\mc F_{\mc A}^*\mu^* g_{\mc A_2}\frac{\mc E_1}{\sqrt 2}\nonumber\\
&\Omega_{\mc B_1}&=&i\lambda_{\mc B}^* g_{\mc B_1}\frac{\mc E_1}{\sqrt 2}+i\mc F_{\mc B}^*\nu g_{\mc B_1}\frac{\mc E_2}{\sqrt 2}\nonumber\\
&\Omega_{\mc B_2}&=&i\xi_{\mc B}^* g_{\mc B_2}\frac{\mc E_2}{\sqrt 2}+i\mc F_{\mc B}^*\nu^* g_{\mc B_2}\frac{\mc E_1}{\sqrt 2}\nonumber\\
&\Gamma_{11}&=&2\cdot(g_{\mc A_1}^2\Im(\lambda_{\mc A})+g_{\mc B_1}^2\Im(\lambda_{\mc B}))+\gamma_1\nonumber\\
&\Gamma_{22}&=&2\cdot(g_{\mc A_2}^2\Im(\xi_{\mc A})+g_{\mc B_2}^2\Im(\xi_{\mc B}))+\gamma_2\nonumber\\
&\Gamma_{12}&=&2\cdot(g_{\mc A_1}g_{\mc A_2}\mu\Im(\mc F_{\mc A})+g_{\mc B_1}g_{\mc B_2}\nu\Im(\mc F_{\mc B}))\nonumber\\
&\Gamma_{21}&=&\Gamma_{12}^*&&\nonumber\\
&\mu&=&(q+p)e^{i\epsilon t}\nonumber\\
&\nu&=&(p-q)e^{i\epsilon t}\nonumber\\
&\lambda_{\mc A}&=&\frac{ f_{\mc A_1}}{1-f_{\mc A_1}f_{\mc A_2}\mu\mu^*}\nonumber\\
&\lambda_{\mc B}&=&\frac{ f_{\mc B_1}}{1-f_{\mc B_1}f_{\mc B_2}\nu\nu^*}\nonumber\\
&\xi_{\mc A}&=&\frac{ f_{\mc A_2}}{1-f_{\mc A_1}f_{\mc A_2}\mu\mu^*}\nonumber\\
&\xi_{\mc B}&=&\frac{ f_{\mc B_2}}{1-f_{\mc B_1}f_{\mc B_2}\nu\nu^*}\nonumber\\
&\mc F_{\mc A}&=&\frac{f_{\mc A_1}f_{\mc A_2}} {1-f_{\mc A_1}f_{\mc A_2}\mu\mu^*}\nonumber\\
&\mc F_{\mc B}&=&\frac{f_{\mc B_1}f_{\mc B_2}} {1-f_{\mc B_1}f_{\mc B_2}\nu\nu^*}\nonumber\\
&f_{\mc A_1}&=&\frac{i}{\kappa_{\mc A_1}-i(\delta_1+h_1)}\nonumber\\
&f_{\mc B_1}&=&\frac{i}{\kappa_{\mc B_1}-i(\delta_1-h_1)}\nonumber\\
&f_{\mc A_2}&=&\frac{i}{\kappa_{\mc A_2}-i(\delta_2+h_2)}\nonumber\\
&f_{\mc B_2}&=&\frac{i}{\kappa_{\mc B_2}-i(\delta_2-h_2)}\nonumber\\
\end{align}

\end{document}